\documentclass{PoS}
\usepackage{xspace}
\usepackage{amsmath}
\usepackage{amssymb}

\newcommand{\aof}{A\,0535+26\xspace}
\newcommand{\hde}{HDE\,245770\xspace}

\newcommand{\herxo}{Her\,X-1\xspace}

\newcommand{\vot}{V\,0332+53\xspace}
\newcommand{\fuoo}{4U\,0115+64\xspace}

\newcommand{\inte}{\textsl{INTEGRAL}\xspace}
\newcommand{\xte}{\textsl{RXTE}\xspace}

\newcommand{\hexte}{\textsl{HEXTE}\xspace}

\newcommand{\swift}{\textsl{Swift}\xspace}

\newcommand{\suzaku}{\textsl{Suzaku}\xspace}

\newcommand{\hexe}{\textsl{HEXE}\xspace}
\newcommand{\osse}{\textsl{OSSE}\xspace}

\newcommand{\ariel}{\textsl{Ariel V}\xspace}

\title{The Be/X-ray binary A0535+26 during its recent 2009/2010 outbursts}

\ShortTitle{The Be/X-ray binary A0535+26 during its recent 2009/2010 outbursts}

\author{\speaker{I.~Caballero}$^a$, 
  K.~Pottschmidt $^b$,
  A.~Santangelo $^c$,
  L.~Barrag\'{a}n $^d$,
  D.~Klochkov $^c$,
  C.~Ferrigno $^e$,
  J.~Rodriguez $^a$, 
  P.~Kretschmar $^f$,
  S.~Suchy $^g$,
  D.~M.~Marcu $^b$,
  D.~M\"{u}ller $^c$,
  J.~Wilms $^d$,
  I.~Kreykenbohm $^d$,
  R.~E.~Rothschild $^g$,
  R.~Staubert $^c$,
  M.~H.~Finger $^h$,
  A.~Camero-Arranz $^{h,i,j}$,
  K.~Makishima $^{k,l}$,
  T.~Mihara $^{l}$,
  M.~Nakajima $^m$,
  T.~Enoto $^k$,
  W.~Iwakiri $^n$,
  Y.~Terada $^n$\\
  \llap{$^a$}  CEA Saclay, DSM/IRFU/SAp --UMR AIM (7158) 
  CNRS/CEA/Universit\'{e} P.~Diderot --F-91191 Gif sur Yvette France\\
  \llap{$^b$} CRESST, UMBC, Baltimore, MD 21250 /NASA
  GSFC, Code 661, Greenbelt, MD 20771, USA\\
\llap{$^c$}Institut f\"{u}r Astronomie und Astrophysik, Sand 1, D-72076 T\"{u}bingen, Germany \\
  \llap{$^d$}  Dr. Karl Remeis-Sternwarte -- FAU Erlangen-N\"{u}rnberg, 96049 Bamberg, Germany\\
  \llap{$^e$} ISDC Data Centre for Astrophysics, 1290 Versoix, Switzerland \\
  \llap{$^f$} ISOC, European Space Astronomy Centre, ESA, 28691 Villanueva de la Ca\~{n}ada, Madrid, Spain\\
  \llap{$^g$}Center for Astrophysics and Space Science, UCSD,  La Jolla, CA, USA\\
  \llap{$^h$}NSSTC, 320 Sparkman Drive NW, Huntsville, AL 35805 USA\\
  \llap{$^i$}Fundaci\'{o}n Espa\~{n}ola de Ciencia y Tecnolog\'{i}a / MICINN,  Madrid, Spain\\
  \llap{$^j$}MICINN (Ministerio de Ciencia e Innovaci\'{o}n), C/Albacete, 5, 28027, Madrid, Espa\~{n}a\\
  \llap{$^k$}University of Tokyo, Japan\\
  \llap{$^l$}RIKEN, Japan\\
  \llap{$^m$}Nihon University, Japan\\
  \llap{$^n$}Saitama University, Japan\\
  E-mail: \email{isabel.caballero@cea.fr}}

\abstract{
The Be/X-ray binary A0535+26 showed a giant outburst in December
2009 that reached $\sim$5.14\,Crab in the 15-50 keV range.
Unfortunately, due to Sun constraints it could not be observed
by most X-ray satellites. The outburst was preceded by four weaker outbursts associated
with the periastron passage of the neutron star. The fourth of them,
in August 2009, presented a peculiar double-peaked light curve, with a first peak lasting about 9 days 
that reached a (15-50 keV) flux of 440\,mCrab. The flux then decreased to less than 220\,mCrab, and increased again reaching 440\,mCrab around the periastron. The outburst was monitored with INTEGRAL, RXTE, and Suzaku TOO observations. One orbital period ($\sim$111 days) after the 2009 giant outburst,
a new and unexpectedly bright outburst took place ($\sim1.4$Crab in the 15-50 keV range). It was monitored with TOO observations with INTEGRAL, RXTE, Suzaku, and Swift. First results of the spectral and timing analysis of these observations are presented, with a specific focus on the cyclotron lines present in the system and its variation with the mass accretion rate.}

\FullConference{8th INTEGRAL Workshop ``The Restless Gamma-ray Universe''\\
   September 27-30 2010\\
    Dublin Castle, Dublin, Ireland}

\begin{document}

\section{Introduction}
\aof is a Be/X-ray binary system, discovered by \ariel in 1975 during a giant outburst \cite{rosenberg75}. 
In this system, a pulsating neutron star ($P_{\mathrm{spin}}\sim103$\,s) orbits around the O9.7IIIe companion \hde 
\cite{giangrande80} in an eccentric orbit of $P_{\mathrm{orb}}\sim111$ days \cite{finger06}. 
An extensive review to the system can be found 
in \cite{giovannelli92}. 
The system is characterized by quiescent states with X-ray luminosity 
$L_{\mathrm{X}}$$\lesssim10^{36}$\,erg\,s$^{-1}$, interrupted by normal (type I) outbursts, with 
$L_{\mathrm{X}}$$\sim$$10^{36-37}$\,erg\,s$^{-1}$ and typically associated 
with the periastron passage, and giant (type II) outbursts, for which 
$L_{\mathrm{X}}>10^{37}$\,erg\,s$^{-1}$. 
Since its discovery, six giant outbursts have been observed. 

After more than 11 years of quiescence the source showed 
a giant outburst in 2005 \cite{tueller05}, which unfortunatelly could not be observed with X-ray 
observatories due to Sun contraints. Two subsequent normal outbursts took place  in 
August/September and December 2005 \cite{finger05}.  
Cyclotron lines were discovered in the X-ray spectrum of the source at 
$\sim$45 and $\sim$100\,keV with \hexe and \osse during giant outbrusts in 1989 and 1994 \cite{kend94,grove95}, and later confirmed with \inte, \xte, and \suzaku \cite{kretschmar05,terada06}. From the cyclotron lines, a magnetic field of B$\sim5.1\times10^{12}$\,G is inferred. 
In some transient X-ray binaries like \fuoo and \vot,  a clear anti-correlation between 
cyclotron line energy and X-ray luminosity has been observed \cite{mihara95,tsygankov06,mowlavi06}.
This anti-correlation has been interpreted as a change of the height 
of the cyclotron scattering region in the neutron star dipole
magnetic field with the accretion rate, assuming the source
is in the super-Eddington regime with a shock front forming 
above the neutron star surface \cite{basko_sunyaev_76}. 
In the case of \herxo, the opposite correlation was discovered \cite{staubert07}.  
The cyclotron energy has been measured to increase with X-ray luminosity, 
suggesting that the source is in the sub-Eddington regime. 
In contrast to these sources, observations of \aof in 2005 with \inte, 
\xte \cite{caballero07}, and  \suzaku \cite{terada06}, show no 
significant variation of the cyclotron line energy, suggesting that 
the line forming region remains stable for different mass accretion rates. 

During the August/September 2005 outburst, a series of short flares  ($\sim1$ day) 
were observed with \swift-BAT during the rise to the peak. From \xte observations of some of these flares, at similar source luminosity to later stages, the cyclotron line appears at a significantly higher position. The pulse profiles also show a different shape and energy dependence than during the rest of 
the outburst \cite{caballero08}. The observed features suggest that 
the flaring activity is caused by magnetospheric instabilities 
on top of quasi-stationary accretion from the disk \cite{postnov08}.

\section{The source in 2009/2010}

\begin{figure}
\centerline{
\includegraphics[angle=90,width=0.6\textwidth]{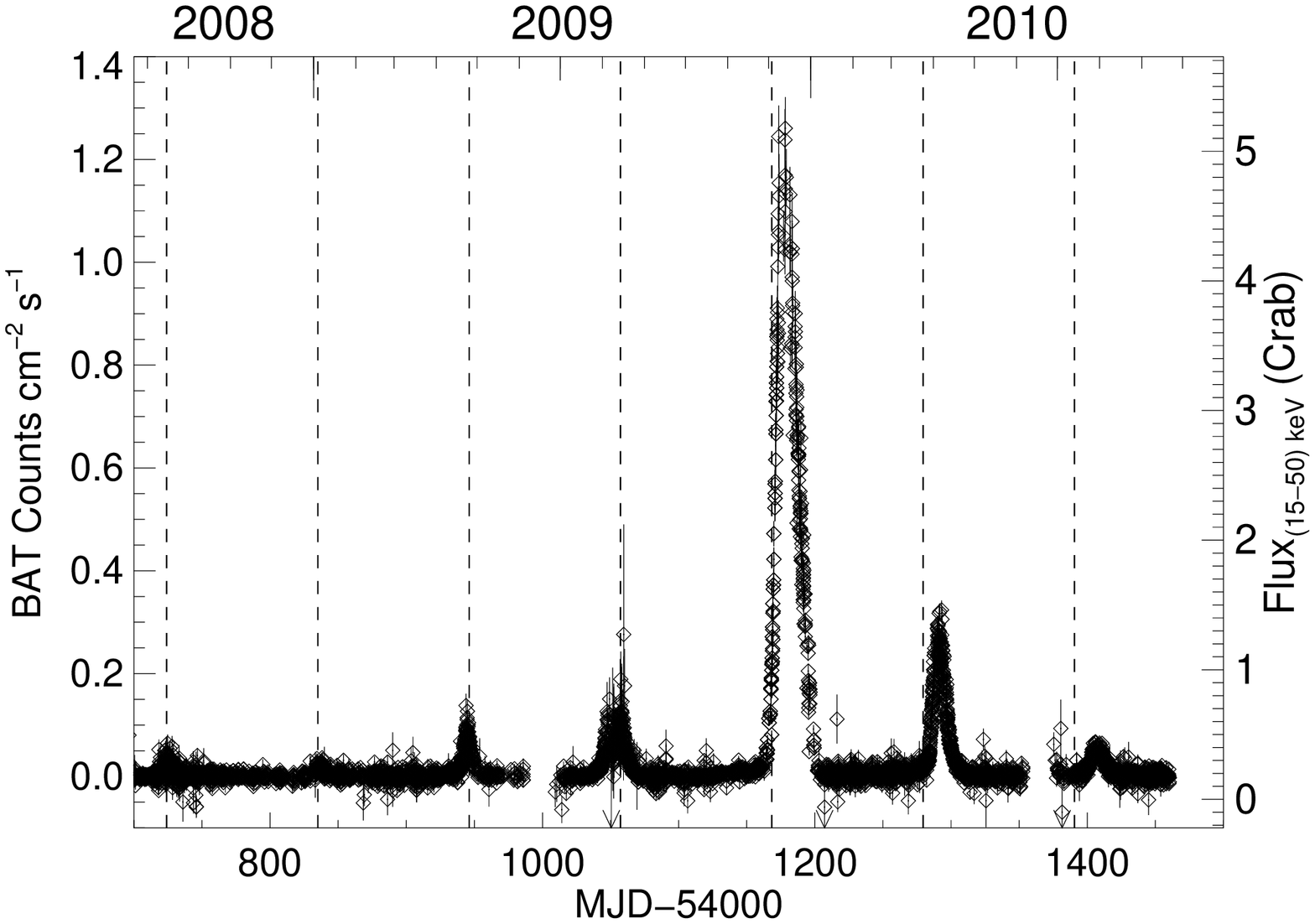}
\includegraphics[angle=90,width=0.6\textwidth]{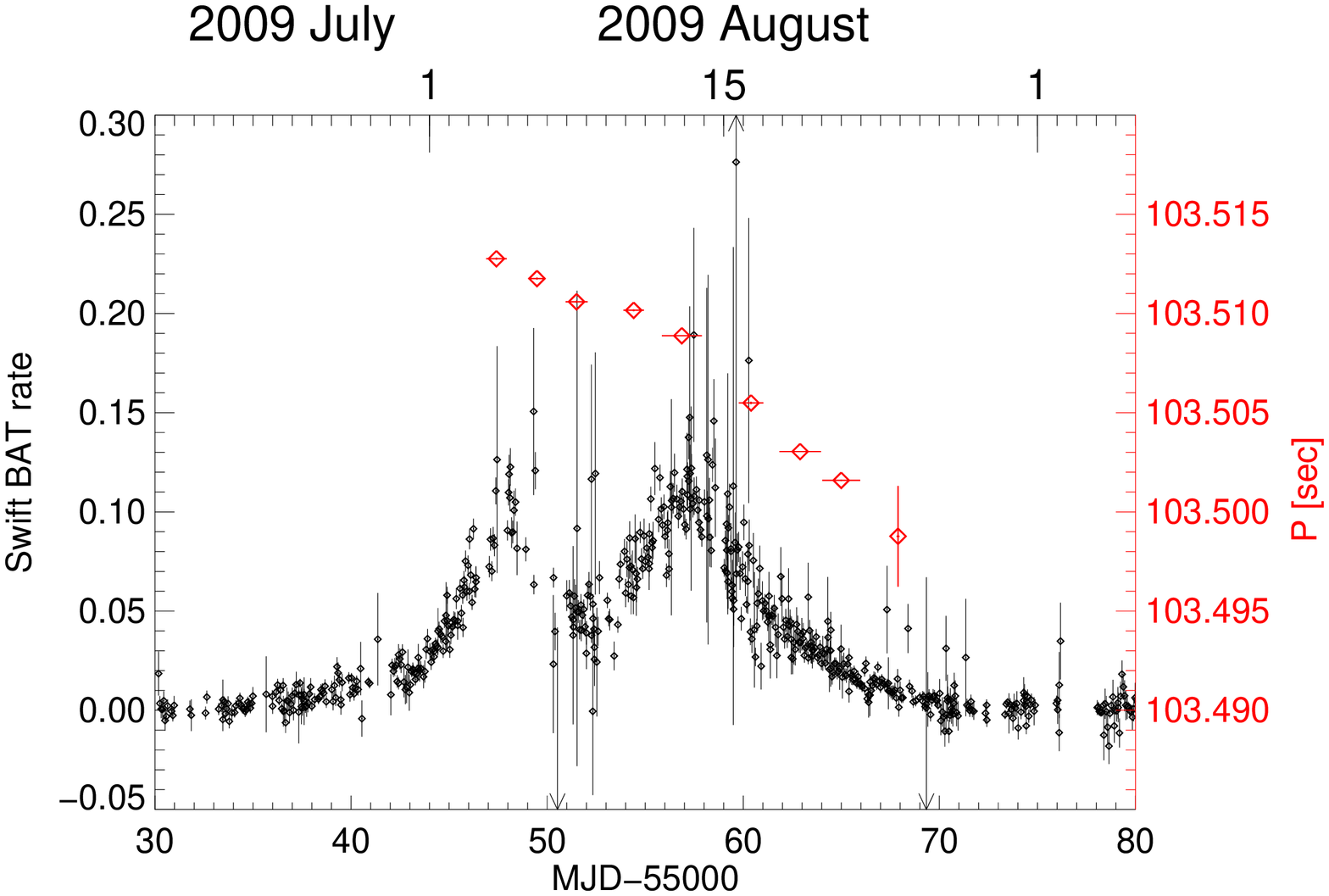}}
\caption{\emph{Left:} \swift-BAT light curve of \aof during its recent 2009/2010 outbursts. The 
vertical dashed lines indicate the times of periastron. \emph{Right:} \swift-BAT light curve
of the double-peaked outburst of \aof in August 2009 (black diamonds, left axis) and pulse period evolution 
measured with \xte data (red diamonds, right axis).}
\label{fig:bat_lc}
\end{figure}

In 2009, the source showed a series of normal outbursts associated with the 
periastron before and after a new giant outburst in December 2009 \cite{finger09,caballero09c}. The giant outburst reached
F$_{(15-50)\,\mathrm{keV}}\sim5.14$\,Crab\footnote{Flux from the \swift-BAT light curve. The Crab flux is about
0.2263 \swift-BAT counts cm$^{-2}$ s$^{-1}$ (15--50\,keV).} (X-ray light curve in  
Fig.~\ref{fig:bat_lc}, left). Unfortunatelly, again due to Sun constraints, the giant outburst could only be observed with 
\xte. The outburst preceding the giant one, that reached a flux F$_{(15-50)\,\mathrm{keV}}\sim440$\,mCrab and presented a peculiar double-peaked light curve (Fig.~\ref{fig:bat_lc}, right), was observed with \inte, \xte, and \suzaku. One orbital period after the giant outburst a new outburst took place in April 2010, unexpectedly bright (F$_{(15-50)\,\mathrm{keV}}\sim1.4$\,Crab) considering it took place after a giant outburst \cite{caballero10}. 

\subsection{Peculiar double-peaked outburst in August 2009} 
The peculiar double-peaked outburst in August 2009 (X-ray light curve in Fig.~\ref{fig:bat_lc}, right) was observed with \inte, \xte, and \suzaku TOO observations. A similar double-peaked outburst was observed with BATSE before
the 1994 giant outburst \cite{finger96}. These pointed TOO observations have allowed us to perform 
a detailed timing and spectral analysis of this peculiar outburst for the first time. 
Using all the \xte-PCA observations, we performed a phase connection method \cite{staubert09} to determine the pulse period evolution. We used the orbital ephemeris from \cite{finger06}. The resulting pulse period evolution is showed in Fig.~\ref{fig:bat_lc} (right). We measure a spin-up of $\dot{P}=-0.5\pm0.04\times10^{-8}\,$s\,s$^{-1}$ (MJD 54046.9). The energy-dependent pulse profiles during the two main peaks of the outburst remain rather stable, and are very similar to pulse profiles observed in past normal and giant outbursts (Caballero et al. 2010, in prep.). 

\begin{figure}
\centerline{
\includegraphics[angle=0,width=0.6\textwidth]{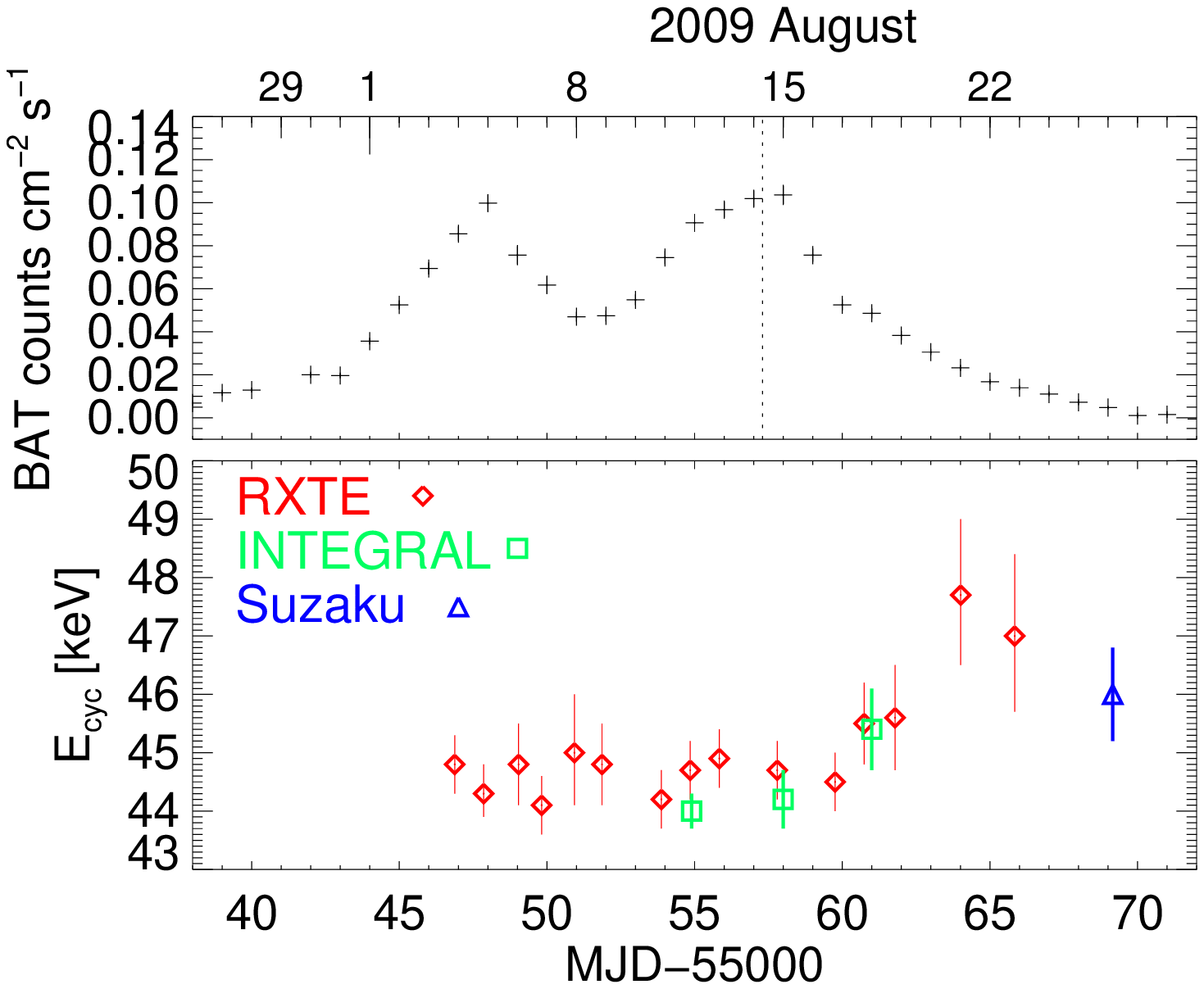}
\includegraphics[angle=0,width=0.6\textwidth]{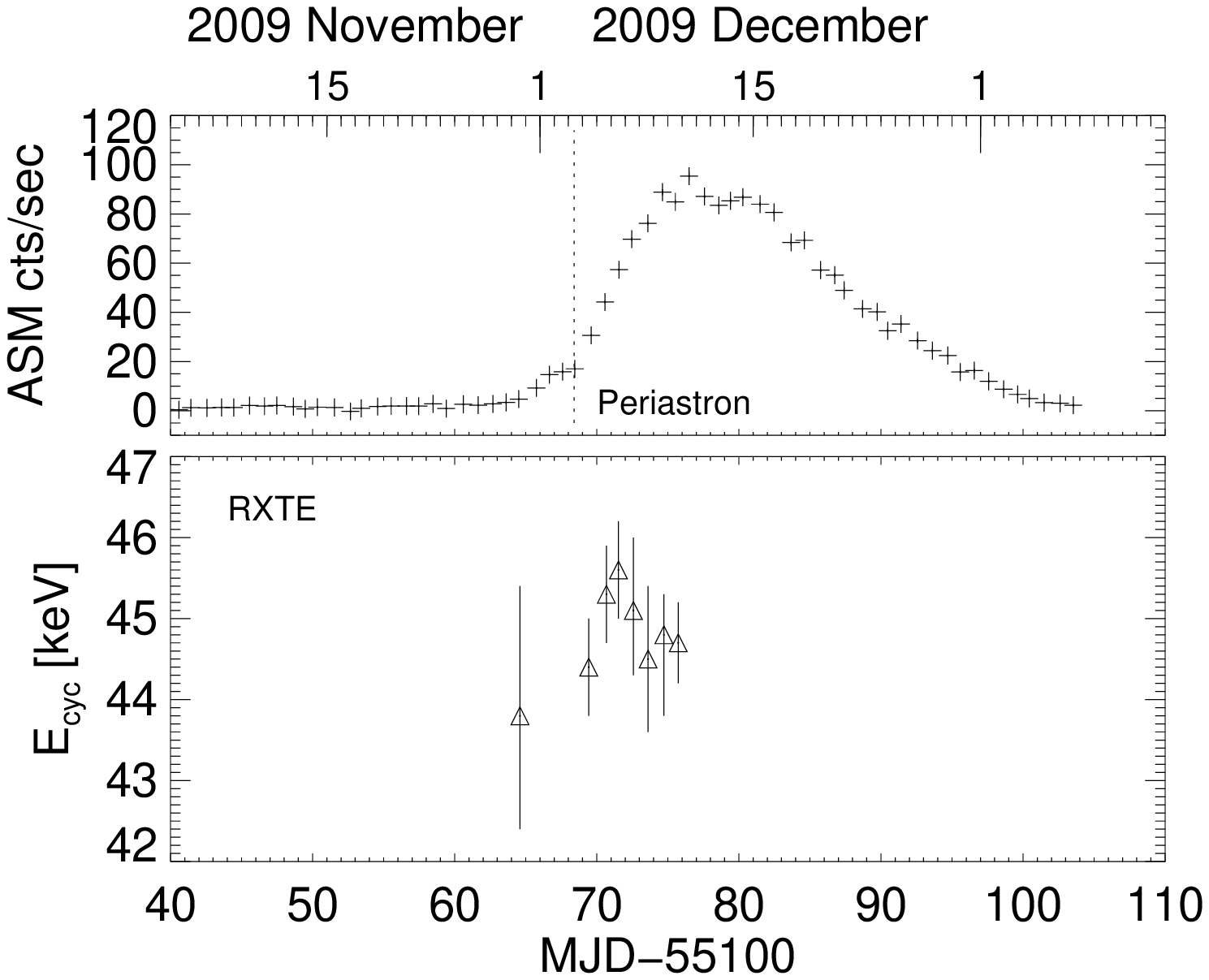}}
\caption{\emph{Left:} \swift-BAT light curve of the August 2009 double-peaked oubturst (top panel), 
evolution of the fundamental cyclotron line energy measured with \inte, \xte and \suzaku (bottom panel). 
\emph{Right}: \xte-ASM light curve of the December 2009 giant outburst (top panel), evolution of the 
fundamental cyclotron line energy measured with \xte (bottom panel) for the observations with \hexte B background
measurements (see text).}
\label{fig:Ecyc}
\end{figure}

Making use of \inte, \xte, and \suzaku, we studied the evolution of the fundamental cyclotron line. We modeled the 
cyclotron lines using Gaussian lines in absorption \cite{coburn02}. 
As shown in Fig.~\ref{fig:Ecyc} (left), the cyclotron line energy remains constant during the two main peaks of the outburst, with a slight increase in the cyclotron line energy during the decay of the outburst. 

\subsection{Giant outburst in December 2009}
The December 2009 outburst was daily monitored with \xte. Unfortunately, at the peak of the outburst the 
HEXTE B cluster stopped rocking, implying that no background measurements are available for the observations 
performed after the peak. Further work is being done to model the HEXTE background properly in order 
to study the cyclotron line position in detail. First results of the cyclotron line evolution during the giant outburst, 
for the observations before the peak, are shown in Fig.~\ref{fig:Ecyc} (right). As can be seen, again no significant variation of the cyclotron line energy with the X-ray luminosity is observed. Work is ongoing to complete this study with all the available \xte observations, that cover the complete outburst. 

\subsection{April 2010 outburst}
\begin{figure}
\centerline{
\includegraphics[angle=0,width=0.6\textwidth]{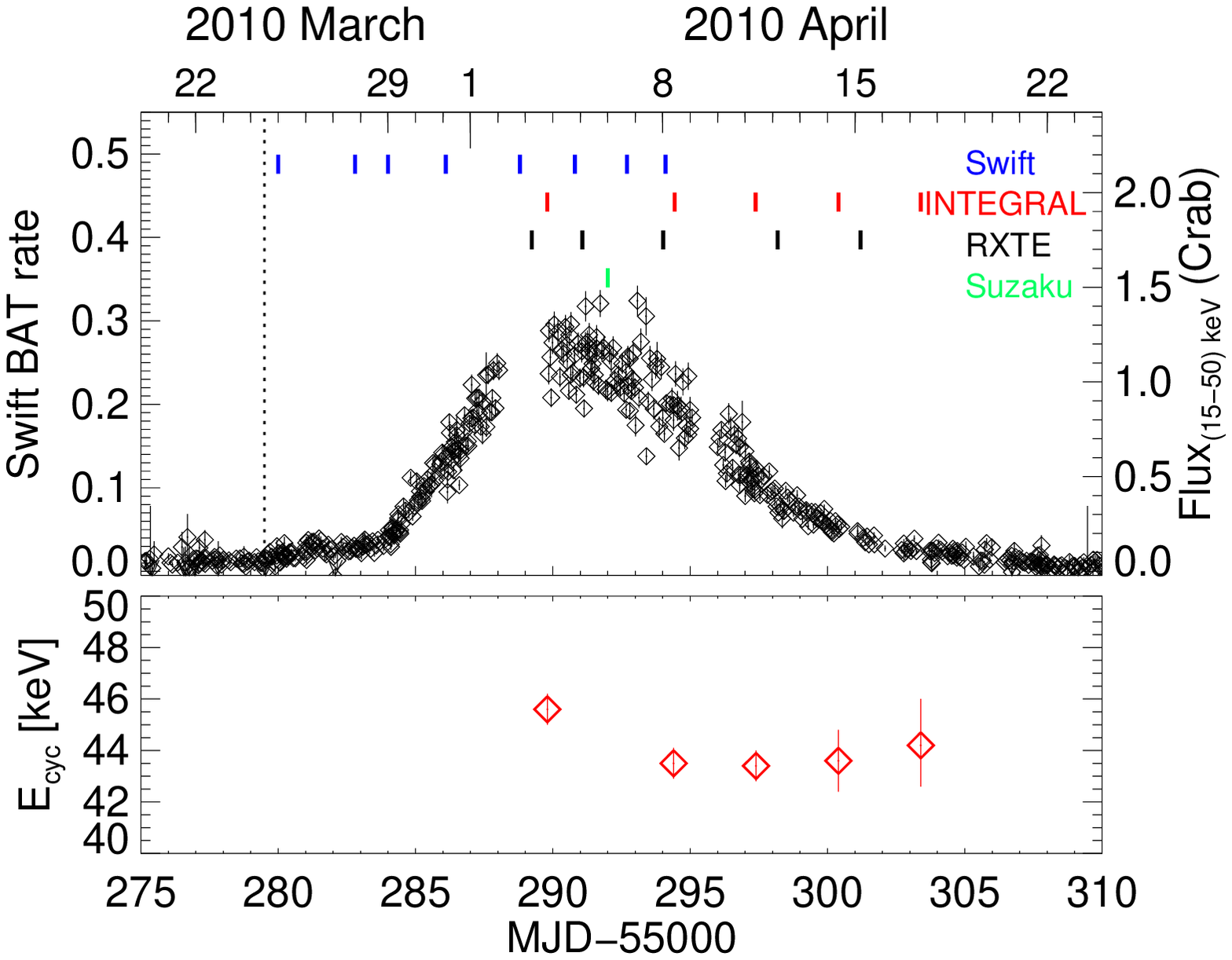}
\includegraphics[angle=0,width=0.6\textwidth]{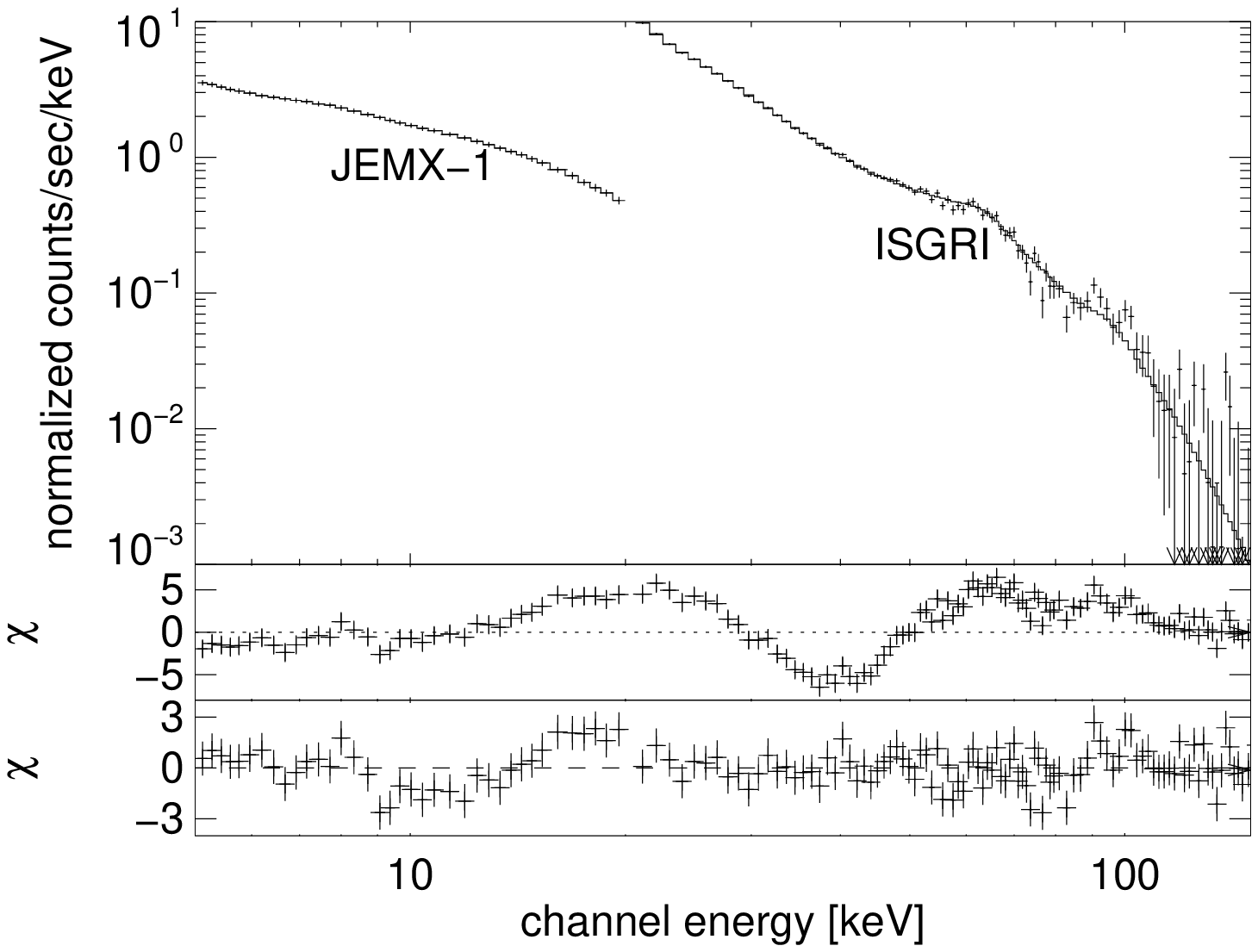}}
\caption{\emph{Left:} \swift-BAT light curve of the April 2010 outburst (top panel). The times of the \inte, \swift, \suzaku, 
and \xte observations are indicated. Bottom panel: evolution of the fundamental cyclotron line energy measured with \inte. 
\emph{Right:} example of \inte spectrum of \aof during the April 2010 outburst (MJD 55297.4). 
The middle panel shows the residuals of a fit without including cyclotron lines in the model. 
The bottom panel shows the residuals of a fit including one cyclotron line at $\sim45$\,keV.}
\label{fig:inte}
\end{figure}

A new outburst took place in April 2010, that reached a flux of F$_{(15-50)\,\mathrm{keV}}\sim1.4$\,Crab. 
Such a bright outburst in principle is not expected after a giant outburst. However, it was predicted from optical observations of the companion star before the outburst, that revealed the presence of a disk of material around 
the Be star from H$\alpha$ lines seen in emission \cite{giovannelli10}. The outburst was observed 
with \inte, \swift, \xte, and \suzaku TOO observations (see Fig.~\ref{fig:inte}, left).  
We studied the evolution of the cyclotron line energy during the outburst. An example of one \inte spectrum 
(MJD 55297.4) is shown in Fig.~\ref{fig:inte} (right).  The presence of the fundamental cyclotron line, measured 
at $E_{\mathrm{cyc}}=43.4\pm0.3$\,keV, is cleary seen in the residuals. 
The evolution of the cyclotron line energy measured with \inte is shown in Fig.~\ref{fig:inte} (left, bottom panel). 

\section{Summary and Outlook}
First results of the recent outbursts of \aof that took place in 2009/2010 have been presented. 
A peculiar double-peaked outburst took place in August 2009 before a giant one 
in December 2009. No significant differences were observed in the timing and spectral analysis of the source compared to past outbursts that could explain the double-peaked shape of the light curve. 
The $H_{\alpha}$ equivalent width shows a dramatic increase between August and December 2009
\cite{camero10}, probing that the disk around the donnor star was growing in that period. The double-peaked light curve
of the August 2009 outburst could be due to the formation of the disk around the Be star. 
The cyclotron line energy evolution remains constant during the two peaks, and a slight increase is measured during the decay of the outburst. 
First results of the cyclotron line evolution during the December 2009 giant outburst and April 2010 outburst  
also reveal no significant variation of the cyclotron line energy. Further work is ongoing to study the spectral and timing properties of the source during these outbursts, making use of all the \inte, \swift, \suzaku, 
and \xte pointed observations. The \swift pointed observations will allow for the first time for this source to study the variation of the $N_{\mathrm{H}}$ during the onset and the peak of the outburst. 

\section*{Acknowledgments}
This work was supported by the Centre National d'Etudes Spatiales (CNES). It is partly based on observations with IBIS embarked on INTEGRAL.

\end{document}